# Extend the FFmpeg Framework to Analyze Media Content

Xintian Wu[1], Pengfei Qu[2], Shaofei Wang[2], Lin Xie[2] and Jie Dong[2]

[1]Intel Corp. Hillsboro, Oregon
[2]Intel China Research Center Ltd, Beijing, China

*Abstract*—This paper introduces a new set of video analytics plugins developed for the FFmpeg framework. Multimedia applications that increasingly utilize the FFmpeg media features for its comprehensive media encoding, decoding, muxing, and demuxing capabilities can now additionally analyze the video content based on AI models. The plugins are thread optimized for best performance overcoming certain FFmpeg threading limitations. The plugins utilize the Intel® OpenVINO™ Toolkit inference engine as the backend. The analytics workloads are accelerated on different platforms such as CPU, GPU, FPGA or specialized analytics accelerators. With our reference implementation, the feature of OpenVINO™ as inference backend has been pushed into FFmpeg mainstream repository. We plan to submit more patches later.

## I. INTRODUCTION

The 21th century is a century of generating information and making use of the information. Information comes from billions of deployed cameras and sensors worldwide to generate business insights; information comes from the massive amount of archived video, audio and text content, waiting to be indexed, made available for search and consumption; and information can also come from social media content that directly or indirectly changes people's life style. Among them, media analytics workloads, such as video summarization and recommendations in video library curation applications, and camera stream analysis in smart city applications, are becoming popular in multiple vertical industries.

Generally, to develop an AI-based applications, a developer first trains a (or a set of) neural network model with any deep learning framework, such as Caffe[2], TensorFlow[8], PyTorch[10], MxNet[11], or Darknet[12]. Next, the developer codes the analytics pipeline and deploys the model in the cloud or on the edge. The pipeline differs with different neural network models, any customized application logics, and whether the pipeline should be accelerated by any platform features or hardware accelerators. There are many tuning parameters that the developer must finetune to archive the optimal performance. It is a big challenge today to develop and deploy a well-optimized AI application.

To accelerate the landscape of AI applications, especially in the case of optimizing analytics workloads on different hardware platforms, industry requires an easy way to build the analytics pipeline and deploy the AI services in the Cloud or at the edge.

Media analytics applications must first be able to process media content. The FFmpeg[1] media framework is an industry de-facto for its comprehensive support on media stream formats, encoding and decoding capabilities, and post-processing features such as resizing and color space conversion. It is a natural choice to extend the FFmpeg framework to support analytics workloads. With FFmpeg analytics, developers can analyze the video (and audio) content in existing or new media applications.

This paper presents a novel way to extend analytics workloads in the existing FFmpeg framework.

Note that recently, as NVidia supports their AI features under a parallel media framework, GStreamer[6]. Many developers start to evaluate the GStreamer framework, which has its pros and cons. The detailed comparison of the two frameworks with regards to running analytics workloads is out of scope of this paper. This paper focuses on enabling the FFmpeg framework.

The rest of the paper is organized as follows: In section 2, we discuss some background context and discuss existing work. In section 3, we propose a novel architecture redesign. In section 4, we present a case study and analyze the performance data. Section 5 provides a summary.

In this article, we use the term "FFmpeg" for the framework and "ffmpeg" (all small capital letters) to refer to the ffmpeg command line tool.

## II. BACKGROUND AND RELATED WORK

### A. Performance Characteristics of Analytics Workloads

In a typical media analytics application, the media is first decoded (to produce media samples) and then the sample content is analyzed to produce the analytics metadata. The performance or structural characteristics of the analytics stage are not the same as the media processing stage. Understanding the difference between the two stages of processing helps us design the right architecture for media analytics applications:

- Unless media encoding is involved, the analytics workloads are generally more computationally intensive than the media decoding and postprocessing tasks. This implies that any optimization effort should first go into optimizing the analytics workloads and then decoding, and/or postprocessing.

- The analytics workloads are usually memory bounded while the media decoding is usually CPU bounded.

- The analytics workloads operate on the RGB color space while the media encoding and decoding favor the YUV space. Therefore, there is usually a post-processing stage in between media decoding and analytics. The post-processing stage converts the color space and resizes the image size. If there are more than 1 analytics model, there could be 1 media decoding followed by multiple independent post-processing and analytics stages in the pipeline, which leads to the importance of enabling threading and data parallelism.

- Threading is critical to archive the overall pipeline performance as the analytics workloads are inherently parallel, where the analysis can be done on multiple streams, multiple image frames, or within multiple regions of an image in parallel. In a typical media analytics pipeline, any upstream or downstream component can stall the pipelining if not properly optimized. Every stage of the pipeline must be optimized for parallelism to archive the overall performance and latency.

- Data format wise, the analytics workloads take tensor (data structures used in neural network models) as input and output. This is usually different from what any media framework typically supports: images and audio frames. Proper abstraction is critical not to confuse developers and increase the media framework learning curve.

- Similarly, in any media framework, the output image/audio data usually goes into an encoder; streams into a URL; or writes to a file. The analytics output instead needs to be stored in a database or sent to a message broker.

As illustrated, extending a media framework to perform analytics workloads is not as straightforward as creating some plugins. The design must consider the pipeline extensibility, the threading efficiency, the I/O capability, the data structure extensibility, and finally the community willingness of extending the media framework to a new territory.

### B. FFmpeg Threading Model

In evaluating extending the FFmpeg framework to analytics workloads, we observe the following limitations in the FFmpeg threading support:

- AVFilter, a library that provides generic audio/video filtering support, is not threaded. This is ok for light workloads. It will become a bottleneck in the pipeline with heavy workloads.

- Filter graph, a directed graph of connected post-processing filters, is not threaded as well. If used in heavy workloads, or with complex post-processing, filter graph will become a bottleneck to optimal performance.

Our work improved the threading support in FFmpeg to overcome these limitations.

### C. FFmpeg Deep Neural Netowrk(DNN) Module

In the FFmpeg mainline, there is a DNN module, which supports some AI workloads. The DNN module can be used for certain simple analytics but it lacks the following features to meet more complex needs:

- Support additional models: The DNN module is hard coded to use TensorFlow as the backend. It is desirable to be able to use other analytics backends.

- Support tensor I/O: The DNN module does not expose tensor data at the input or output, which makes it difficult to chain additional analytics operations, such as object tracking, after the DNN module (object detection.)

- Customize AI filters: There is limited processing filters that can be used with the DNN module. It is desirable to be able to customize the pre-processing filters.

Our work redesigned the analytics support in FFmpeg and can support most of analytics workloads. With our reference implementation, the feature of using OpenVINO™ as inference backend has been pushed into FFmpeg mainstream repository. Some more features for OpenVINO™ are being added into the DNN module, most of ideas are coming from our implementation.

### D. Other Media Frameworks

In addition to FFmpeg, there are other media frameworks that can be extended to perform analytics work. They are listed here for reference:

- DL Streamer: Deep Learning (DL) Streamer [5] contains the GStreamer plugins that enable CNN model-based video analytics capabilities, utilizing the Intel® OpenVINO™ Inference Engine. This repository implements similar work as ours, except that the support is for the GStreamer media framework.

- MediaPipe: With MediaPipe[3][7], a perception pipeline can be built as a graph of modular components, including, for instance, inference models and media processing functions. MediaPipe depends on FFmpeg to handle media processing such as decoding and demuxing media streams.

- G-API: OpenCV 4.0 introduced the Graph API (G-API)[4], which specifies a sequence of OpenCV image processing operations in the form of a graph. In comparison, the FFmpeg framework supports operations on arbitrary data types and has native support for streaming time-series data, which makes it much more suitable for analyzing audio and video data.
- Deep Stream SDK: It is a closed source SDK and based on the GStreamer Framework. The user can build streaming pipelines for AI-based video and image analytics using Deep Stream SDK.

### III. ARCHITECHTURE

#### A. Design

The new set of analytics plugins are designed as FFmpeg AV filters. The extensibility of AV filters allows us to easily implement different AI pre-processing filters. The ability to connect multiple AV filters in a graph helps to construct complex AI pipelines, with the convenience of executing the filter graphic right in the ffmpeg command line tool. Figure 1 shows the overall design:

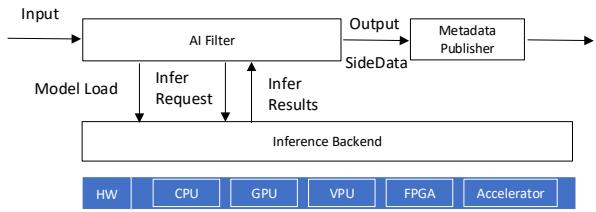

Fig. 1. Architecture Design

The new design utilizes the AVFilter extension and implements object detection, classification, and recognition as a set of AVFilter plugins. The filters can work as a chain of operations in a filter graph.

A typical analytics pipeline consists of a demuxer, a decoder, a post-processing filter, and a few analytics filters. With the new design, it is easy to extend the pipeline with new analytics features.

#### B. Metadata

Metadata stores analytics related information in the data structure AVFrameSideData. Metadata is attached as the side data in the frame info structure AVFrame, transferred from filter to filter. Metadata can be converted to a standard format by the metadata publisher.

As analytics output usually requires some post-processing, it is desired to unify the way postprocessing is handled. Each AI filter has an attribute named "model_proc", which can be used to specify the postprocessing requirement of the inference output. There is no need to specifically code postprocessing into the pipeline after specifying "model_proc".

#### C. AI Filter

Based on the architecture design, we implement two AI filters to demonstrate how to build the pipeline.

- The "detect" filter does object detection only. It takes video frames as input from an upstream filter and performs the inference operation on it. The detection results are a list of bounding boxes and are stored in the side data of each frame structure. Besides the coordinates of the bounding box, label id and the confidence of the detection result are also stored. After the inference, it passes the result to a downstream filter.
- The "classify" filter is designed to be inserted into the pipeline after a "detect" filter. Multiple "classify" filters can be inserted, to run different classification models. The "classify" filter executes the inference operation on the objects detected from the "detect" filter. There may be zero or multiple objects detected. For each detected object, the results are cropped, scaled and converted to the BGRP format (the BGRP format generally is supported by the AI models), and finally, the inference operation is executed.

The "detect" and "classify" filters are designed in a similar way. The neural network model with the same input and output can be fed into the filter. It is easy to switch models and quickly verify their operations in the pipeline.

#### D. Metadata Publisher

The metadata publisher receives the meta data from an upstream filter and converts it in a human readable format such as json, text, or Kafka message.

The metadata publisher can send the output to a Kafka message broker, which is convenient for applications that need to exchange metadata with other services running in a cloud setup.

#### E. Inference Backend

The Inference Backend is a bridge layer between the AI filters and an inference engine implementation. The backend implements resource management and passes the inference request to an inference engine implementation. The backend is transparent to the developers.

- The reference implementation utilizes the Intel® OpenVINO™ Toolkit Inference Engine as a backend implementation. Intel® OpenVINO™ Toolkit is a high-performance inference framework optimized for Intel CPU, GPU, VPU, FPGA and other analytics accelerators. With this backend implementation, an AI application can be deployed to a variety of Intel platforms. This feature has been pushed into FFmpeg mainstream repository.
- Besides Intel® OpenVINO™ Toolkit Inference Engine, the backend can use any other inference engines, such as the Tensor Flow inference engine, as along as a set of standardized inference interfaces is implemented.

#### F. Threading Optimization

We have optimized the FFmpeg framework threading model at the frame level and at the component (decoder or filter) level. The optimization minimized the impact of threading issues described in section 2.

Within a filter, threading is optimized at the frame level. Inference requests on multiple objects can run in parallel to increase the inference throughput.

Threading is also optimized between the filters at the filter graph level. Each filter is optimized to run in a separate thread.

## IV. CASE STUDY

Two use cases are described in this section. Both use cases are based on the FFVA reference implementation. The first use case shows how to construct a simple analytics pipeline. The second use case involves a more complex pipeline topology.

In both use cases, we measure the performance data to show the performance difference with or without enabling threading. In the non-threaded mode, each component runs in serial. In the threaded mode, each component runs in its own thread, including the decoder and any post-processing filters. Note that the threading optimization requires non-trivial changes to the FFmpeg code. It is in the early experimental stage and is not yet released into the FFVA repository. We show the data here to illustrate what's possible with FFmpeg.

Fig. 2 and Fig. 3 show the hardware and software configurations used to collect the performance data:

| Component | Description |
| --- | --- |
| Platform | E5 Server with Intel® Xeon® code named Cascade Lake 6252 CPU |
| Memory | 40G |
| OS | Ubuntu 18.04.2 LTS |
| FFVA version | Release 0.4.2 |
| FFmpeg version | Release 4.2 |
| Experimental Features to support threading and parallel mode | Experiment only. It is in private repo and not part of the FFmpeg or FFVA |

Fig. 2. Hardware and Software Configuration

| Case | Model |
| --- | --- |
| Face Detection | face-detection-adas-0001 |
| Face Recognition | face-reidentification-retail-0095 |
| Car/Plate Detection | vehicle-license-plate-detection-barrier-0106 |
| Plate Recognition | license-plate-recognition-barrier-0001 |

Fig. 3. Test Models from Intel Open Model Zoo[9]

### A. Car Detection and License Plate Recognition

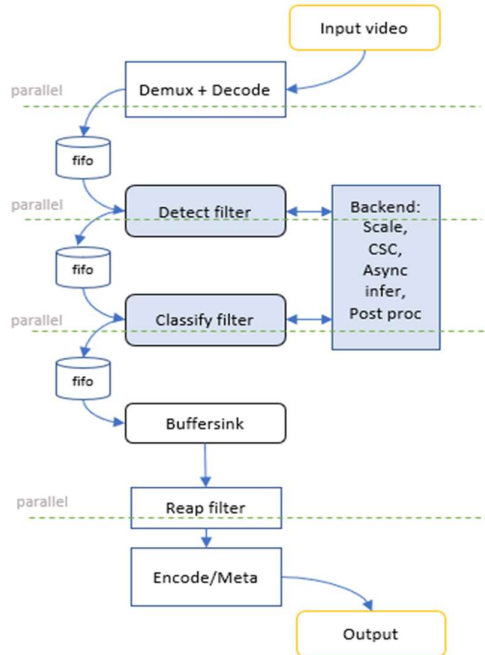

Fig. 4. The Pipeline of Car Detection and License Plate Recognition. Based on the FFmpeg framework, the main components in the pipeline can be paralleled in their dedicated thread. The image scaling, color space conversion, asyncronized inference and part of post processing can be handled by the Backend.

In this use case, the analytics pipeline includes a demuxer, a decoder, a color space convertor, a scaling filter and two AI inference filters: the "detect" filter, which detects cars and license plates, and the "classify" filter, which recognizes the plate letters and numbers.

The command to build and run the pipeline is as follows:

$ffmpeg -i xxx.mp4 -vf

detect=model=vehicle-license-plate-detection-barrier-0106:device=CPU:nireq=$NIREQ:configs=CPU_THROUGHPUT_STREAMS=24\,CPU_THREADS_NUM=96,

classify=model=license-plate-recognition-barrier-0001:device=CPU:nireq=$NIREQ:configs=CPU_THROUGHPUT_STREAMS=24\,CPU_THREADS_NUM=96

-y -f null -

With the same command and parameters, the E2E performance is showed as the following:

| Mode | E2E Performance |
| --- | --- |
| Parallel mode | 777 fps |
| Non-Parallel mode | 653 fps |

Fig. 5. E2E Performance of Case A

The above table shows about 19% performance improvement compared to the non-threaded mode.

The reason why the threaded mode archived better performance is due to the improved resource utilization, including CPU and memory. Each pipeline component, the decoder, the detection filter and the classification filter, works in parallel.

The "-nireq" parameter specifies the maximum number of inference requests that can be processed simultaneously by the AI filter. A higher "-nireq" value can result in better parallelism thus better performance, at the same time consuming more system resources.

Some analysis of the detailed pipeline execution is as follows: At the beginning, all inference request resource is in an idle state as there's no inference task. A few of them become busy when an inference request comes in. After the inference task is finished, the state of the inference resource is changed back to the 'idle' state.

As inferencing is usually the dominant workload within a pipeline, the more inference requests get served (status busy), the better the pipeline throughput, as well as the more efficient the pipeline execution scheduling. For the latter, the inference request busy statistics can be impacted if the pipeline execution is stalled by any slower pipeline component, for example, if decoding is slow to producing video frames. Efficient scheduling makes sure that all stages of the pipeline run in parallel and there is no stalled component (due to upstream or downstream bottlenecks.) Fig. 6 shows the inference resource utilization between the non-threaded and threaded modes. The threaded version shows a better inference resource utilization:

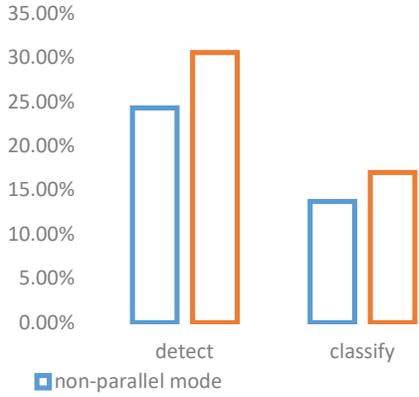

Fig. 6.  Ratio of Resource Utilization

*B.  1: N With Face or License Plate Recognition*

This use case adds multiple face and license plate recognitions into the pipeline. The pipeline includes two branches. Each branch is composed of two AI filters for detection and classification. The first branch performs car detection and license plate recognition. The second branch works on face detection and face recognition. Both branches analyze the same input video sequence.

Again, in the non-threaded mode, and the AI filters run in serial. In the threaded mode, the AI filters run in their own threads.

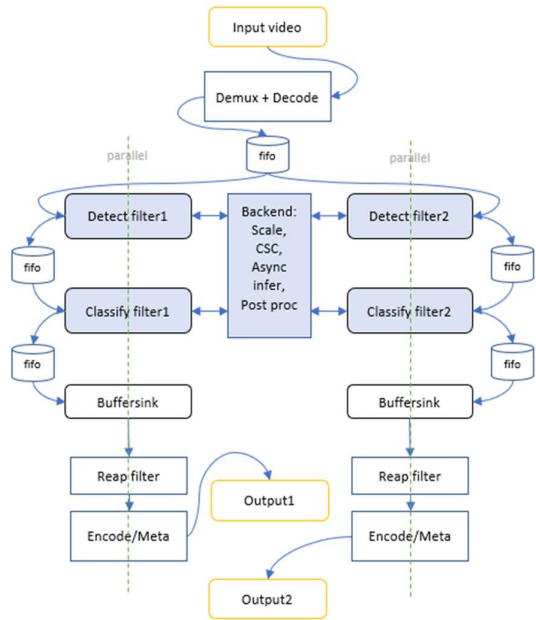

Fig. 7.  The Pipeline of 1: 2 with Face Recognition and License Plate Recognition

The ffmpeg command line that constructs the analytics pipeline is as follows:

$ffmpeg -i xxx.mp4 -vf

detect=model=vehicle-license-plate-detection-barrier-0106:device=CPU:nireq=$NIREQ:configs=CPU_THROUGHPUT_STREAMS=24\,CPU_THREADS_NUM=96,

classify=model=license-plate-recognition-barrier-0001:device=CPU:nireq=$NIREQ:configs=CPU_THROUGHPUT_STREAMS=24\,CPU_THREADS_NUM=96

-y -f null - -vf

detect=model=face-detection-adas-0001:device=CPU:nireq=$NIREQ:configs=CPU_THROUGHPUT_STREAMS=24\,CPU_THREADS_NUM=96,

classify=model=face-reidentification-retail-0095:device=CPU:nireq=$NIREQ:configs=CPU_THROUGHPUT_STREAMS=24\,CPU_THREADS_NUM=96

-an -y -f null - -abr_pipeline

Note that the "-abr_pipeline" option is a feature implemented in FFVA to enable parallelism of the ffmpeg filter graph (the components chain in the green dash line can be threaded in "Fig. 7") for a specific case of running a single input and multiple filter graph outputs.

The performance data is shown in Fig 8:

| Mode | E2E Performance |
|---|---|
| Parallel mode | 307 fps |
| Non-Parallel mode | 245 fps |

Fig. 8.  E2E Performance of Case B

The table shows about 25% performance improvement compared to the non-threaded mode.

The new design overcomes some of the threading limitation in the FFmpeg framework. It enables threading in the filter graph and demonstrates the design and scheduling effectiveness.

## V.  CONCLUSION

In this paper, we presented FFVA, an extension of the FFmpeg framework to support analytics workloads. We described the FFVA architecture with the reference implementation. The two typical use cases showed that simple or complex analytics pipelines can be equally easy to construct with the FFVA extension. We further showed what could be additionally improved in FFmpeg in terms of threading and measured the performance data.

The key contributions of this paper are (a) expanded the FFmpeg framework to program simple or complex AI related use cases; (b) leveraged the Intel® OpenVINO™ Toolkit Inference Engine to deploy any AI applications on a wide range of hardware platforms; (c) explored the performance optimization opportunities in the FFmpeg framework.

ACKNOWLEDGMENT

We would like to express our deepest gratitude to all for their advice and guidance in our works.

REFERENCES

[1]  FFmpeg http://ffmpeg.org/documentation.html
[2]  Caffe2. https://caffe2.ai. Last accessed on 2019- 04-12
[3]  Camillo Lugaresi, Jiuqiang Tang, Hadon Nash, Chris McClanahan, Esha Uboweja, Michael Hays, Fan Zhang, Chuo-Ling Chang, Ming Guang Yong, Juhyun Lee, Wan-Teh Chang, Wei

Hua, Manfred Georg, Matthias Grundmann, "MediaPipe: A Framework for Building Perception Pipelines", arXiv:1906.08172

[4] Dmitry Matveev. OpenCV Graph API. Intel Corporation, 2018.
[5] DL Streamer, https://github.com/openvinotoolkit/dlstreamer_gst
[6] GStreamer. The GStreamer Library, 2001. https://gstreamer.freedesktop.org/ .
[7] MediaPipe Google. https://github.com/google/mediapipe
[8] TensorFlow. TensorFlow Lite, 2017. https://www.tensorflow.org/lite , Last accessed on 2019-04-11.
[9] Intel Open Model Zoo, https://github.com/opencv/open_model_zoo
[10] Pytorch, https://pytorch.org/
[11] Apache MXNet, https://mxnet.incubator.apache.org/
[12] Darknet, https://pjreddie.com/darknet/